\documentstyle[aps,prl,twocolumn,psfig]{revtex}
\begin{document}
\draft                       

\title{Characterization of the channel diffusion in a sodium 
tetrasilicate glass via molecular-dynamics simulations}
\author{Emmanuel Sunyer, Philippe Jund$^*$  and R\'emi Jullien}
\address{Laboratoire des Verres, Universit\'e Montpellier 2, Place 
E. Bataillon, Case 069,\\34095 Montpellier, France\\
$^*$ Laboratoire de Physicochimie de la Mati\`ere Condens\'ee, 
CNRS-Universit\'e Montpellier 2,\\Place E. Bataillon Case 03, 
34095 Montpellier, France}

\maketitle

\begin{abstract}
We study the structural and dynamical characteristics of the sodium atoms 
inside and outside the ``diffusion channels'' in glassy Na$_2$O-4SiO$_2$ 
(NS4) using classical molecular dynamics. We show that on average neither
energetic arguments nor local environment considerations can explain
the increased density of sodium atoms inside the subspace made of the
channels. Nevertheless we show that at low temperature the mean 
square displacement of the sodium atoms inside this subspace is 
significantly larger than the one of the atoms outside the channels. 
\end{abstract}

\pacs{PACS numbers: 61.20.Ja, 61.43.Fs, 66.30.Hs} 

\narrowtext 

The diffusion of alkali atoms in silicate glasses is an important matter under
investigation since several years. In particular the properties of 
sodium atoms inside the amorphous tetrahedral network of silica have been
the topic of both experimental work ~\cite{exp} and molecular-dynamics 
simulations (MD)~\cite{MD} since a simple glass like Na$_2$O-4SiO$_2$ (NS4) 
can be used as a prototype for more complicated glasses. The question of
how the alkali atoms diffuse inside the glassy network is still a matter of
debate and a generally accepted theory of ion transport in glasses is
still missing \cite{theories,roling}. In a previous MD study of NS4 we 
have shown that the ions follow preferential pathways (``channels'') inside
the glassy matrix \cite{Jund}. Nevertheless contrarily to the popular
idea proposed by Greaves \cite{Greaves} and developed in further studies 
\cite{seg}, these channels are
neither static nor due to a microsegregation of the sodium atoms 
but have to be seen
dynamically in the sense that the channels are those regions of space in
which a great number of sodiums have {\em passed} during a given simulation 
time. The existence of these channels gives rise to a pre-peak in the
structure factor at around
$q=0.95$~\AA$^{-1}$
seen both in experiments \cite{Meyer} and classical MD simulations
~\cite{horbach}.
In this last paper Horbach
\textit{et al.}
show that the slow dynamics of the sodium atoms is closely 
related to the one of the underlying silica matrix which is coherent
with the fact that there exists a strong correlation between the 
channels and the location of the non-bridging oxygen atoms, as shown
in previous MD studies
~\cite{Oviedo,Sunyer}.\\
Once the existence of the channels is established, the next step is
naturally to analyze their characteristics and to determine {\em why} the
sodium atoms take these
preferential
pathways. This is the aim of the present study. In that direction we analyze
the potential energy and the local structure of the sodium atoms whether they
are INside the channels (Na$_{\rm In}$) or OUTside the channels (Na$_{\rm Out}$).
It is indeed generally believed that the diffusion of the ions occurs via
``hopping motions between well-defined potential minima'' \cite{roling} which
we should be able to detect in our simulations. We analyze also the time
evolution of the sodium densities inside and outside the channels which permits
us to detect the dynamics of the channels as a function of temperature and 
to quantify the differences between the sodium densities, differences that 
so far have only been suggested \cite{horbach}. To elucidate the diffusion
mechanisms of the Na atoms we determine a characteristic ``residence'' 
time inside the channels and we study the mean square displacements versus 
temperature of the Na$_{\rm In}$ and Na$_{\rm Out}$ atoms.

Our system is made of 648 particles, namely $N_{\rm{Na}}=$~86 sodium, 173 
silicon and 389 oxygen atoms, confined in a cubic simulation box of edge 
length $L=20.88$ \AA\ to which we apply periodic boundary conditions. The 
density is thus the experimental one, i.e. 2.38~g~cm$^{-3}$~\cite{Doremus}. 
The interaction potential used in our simulations is a modified version 
of the one proposed by Kramer \textit{et al}~\cite{BKS} and its complete 
description can be found in~\cite{Sunyer}. It has been shown in 
previous studies that 
this potential is able to describe reliably many structural and dynamical 
properties of different sodium silicate melts~\cite{horbach,Horbach} 
and especially NS4~\cite{Jund,Sunyer}. We have used as initial structure 
a $\beta$-cristobalite crystal in which we have randomly substituted the 
appropriate number of SiO$_4$ tetrahedra by Na$_2$O$_3$ ``molecules''. 
Subsequently the system is melted and equilibrated in the liquid phase 
at high temperature, i.e. 4000~K, for 50000 time steps ($\equiv 35$ ps). 
Then it is cooled down very rapidly with a linear cooling
schedule at a quench rate of $2.3 \times 10^{14}$ K~s$^{-1}$. During the 
quenching process, the configurations of the system (positions and velocities
of all the atoms) are saved at different temperatures ($T \approx$ 4000, 3100,
2500, 2300, 1900 and 1700~K). These configurations are subsequently used as 
starting points of production runs of 2 million steps ($\equiv 2.8$ ns) 
performed in the micro-canonical ensemble [($N$,~$V$,~$E$)~$=$~const]. During
these production runs we save 2000 configurations equally spaced in time.
The glass transition temperature of the system can be roughly estimated from
the location of the bend in a plot of the potential energy vs $T$ to be around 
$T_g \approx$~2400~K. At each temperature the results have been averaged over three independent
samples.

As mentioned above, we have shown in previous studies~\cite{Jund,Sunyer} that
at low temperature the sodium atoms diffuse through the (quasi-)frozen silica 
matrix within a subset of the total available space. This subset is made of 
pockets (typical size 3-6~\AA) connected together via small pathways (similar
in a sense to a neuronal network) and this network is what we call channels.
Our aim in this work is to characterize the channel diffusion of the sodium atoms, 
i.e. to study the possible structural and dynamical differences between the Na$_{\rm In}$
and the Na$_{\rm Out}$ atoms. In order to determine the channels,
we divide the simulation box in $N_{tot}=20^3$ small distinct cubes of volume 
$\approx$~1~\AA$^3$ and determine the number density of the Na atoms in each
of these cubes during the 2.8 ns of the simulation. Then we consider only the 
upper 10~\%
of all the visited cubes. These cubes form ``the core'' of the 
channels (the space in which the sodium atoms have been the more often) or 
shortly ``the channels''. More details can be found in Ref.~\cite{Jund,Sunyer}. 
Once the channels have been defined via this time integration, we can go
back and determine for every time $t$, Na$_{\rm In}$ and Na$_{\rm Out}$ and start 
our investigations.\\
Why do the Na atoms visit a specific fraction of the total available 
space more frequently at low temperature? The first explanation that comes 
to mind is that the visited sites are more favorable energetically. To check 
this idea, we have calculated the individual potential energy distributions 
of the sodium atoms at 
1700~K
whether they are inside or outside the channels. The two distributions have
been averaged over the 2000 configurations saved at
1700~K
and over the 3 independent samples and are represented in \textsc{Fig.}~1.
It is obvious in
\textsc{Fig.}~1
that there is no significant difference between the distribution of 
Na$_{\rm In}$ and Na$_{\rm Out}$: both are gaussians centered approximately
at $-$3.8~eV with a Half Width at Half Maximum of $\approx$ 0.6~eV.
It appears therefore that the energy is not the driving force explaining why
the Na atoms are inside the channels. One may argue that the time average can
smear out the effect but in our previous study we have shown that the 
sodium atoms visit sites previously occupied by other sodium atoms \cite{Jund} and therefore if 
these sites were favorable energetically it should appear in
\textsc{Fig.}~1. We have also calculated the energy distributions of Si 
and O atoms (not shown in \textsc{Fig.}~1
 ) and found that they have an average energy of
$-$35~eV and $-$12~eV respectively.\\ 
In fact, the individual potential energy distributions reflect the whole 
environment of the Na atoms. Therefore even if there exists a structural
difference at short range, it will not be visible in
\textsc{Fig}.~1 because of all the other (long-range) interactions. 
It is hence justified to look more precisely
at the environment of the sodium atoms via the radial pair distribution 
function
and the integrated number of neighbors. For a given $\alpha - \beta$ pair
they are defined by:
\begin{eqnarray}
g(r)_{\alpha - \beta}=\frac{V}{4 \pi r^{2} N_{\alpha} dr}~dn_{\beta}
\end{eqnarray}
and,
\begin{eqnarray}
N(r)_{\alpha - \beta}=\frac{N_{\alpha}}{V}\int_{0}^{r}4 \pi r^{2}
 g(r)_{\alpha - \beta}~dr
\end{eqnarray}
Note that the number of first nearest $\beta$ neighbors around species
$\alpha$, is given by the value of $N(r)_{\alpha - \beta}$ at the first minimum
of $g_{\alpha - \beta}(r)$. In \textsc{Fig.}~2 we have represented $g(r)$ and
$N(r)$ at $T$~=~1700~K for~(a)~Na$_{\rm Out}-$Na, Na$_{\rm In}-$Na and Na$-$Na
pairs,~(b)~Na$_{\rm Out}-$O, Na$_{\rm In}-$O and Na$-$O pairs 
and~(c)~Na$_{\rm Out}-$Si, Na$_{\rm In}-$Si and Na$-$Si pairs. The major
differences can be seen in (a) where it appears that a Na$_{\rm In}$ has on
average 1 supplemental Na neighbor
compared to a Na$_{\rm Out}$ atom. This goes together with a reduction of the
number of nearest Si neighbors~(c).
On the contrary the local oxygen environment~(b)
seems to be the same for the sodiums inside and outside the channels.
Finally concerning the nearest
neighbor distances with the other species there is no significant 
change between Na$_{\rm In}$ and  Na$_{\rm Out}$ except a slight decrease
(0.1~\AA) of the Na$_{\rm In}-$Na distance compared to the distance between
Na$_{\rm Out}$ and the other sodium atoms. It appears therefore that the local
structure of the sodium atoms inside and outside the channels is not 
significantly different in order to explain the existence of the channels.
Nevertheless it appears from the study of the radial pair distribution
functions that on average the Na$_{\rm In}$ atoms have more Na neighbors
than the Na$_{\rm Out}$ atoms. Therefore it is justified to calculate at any
instant $t$ the number of sodiums inside the channels
($N^{\rm In}_{\rm Na}(t)$) and the number of sodiums outside the channels
($N^{\rm Out}_{\rm Na}(t)$). These quantities are shown in \textsc{Fig.}~3
at 1700~K~(a),~2300~K~(b) and 3100~K~(c).
A first information can be obtained from the shape of the curves at the
different temperatures. At 1700~K~(a), the fact that the curves
become flat after $\approx 0.5$~ns
indicates that the channels are well defined through the frozen silica matrix
and are made almost of the same small cubes during the rest of the
simulation. At 2300~K~(b), the situation is rather different since
$N^{\rm In}_{\rm Na}$ goes through a maximum (and consequently 
$N^{\rm Out}_{\rm Na}$ goes through a minimum) after
$\approx 1.4$~ns.
This is a signature of the residual motion of the silica matrix which is
only {\em quasi}-frozen and hence a manifestation of the motion of 
the channels.
In fact, the channels that we have defined in our calculation at that
temperature are the common part of the ``moving'' channels we could have
defined over consecutive fractions of the whole simulation. It is therefore
reasonable to assume that the real channels at
$\approx 1.4$~ns (half the simulation length) are closest to the 
channels that we have defined
over the whole simulation. At 3100~K~(c), the fact that the curves are flat
indicates that the channels do not really exist anymore. What we have defined
as channels at that temperature are only randomly distributed clusters of
density fluctuations. Actually, the matrix is melted and the sodium atoms may
visit the entire accessible space (the system is ergodic) and hence the
location of the clusters is completely random and depends crucially on the
length of the run. Thus, the definition of the channels is quite artificial
at high temperature. Nevertheless when we compare the values of
$N^{\rm In}_{\rm Na}$ after
$\approx 1.4$~ns for the different temperatures we do not see an increase of
$N^{\rm In}_{\rm Na}$ with increasing temperature contrarily to what was 
suggested by Meyer \textit{et al.}~\cite{Meyer}
to explain an increase of the quasielastic amplitude in the dynamic structure
factor of sodium disilicate [NS2] melts. Finally we see in
\textsc{Fig.}~3
that for all the temperatures the major part of the sodium atoms is outside the
channels. Nevertheless as shown
in~\cite{Jund} the total volume of the channels is small
($\approx$~10~\% at 2000~K) therefore the sodium densities can be 
very different 
inside and outside the channels. In \textsc{Fig.}~4
we have represented the sodium density $\rho_{\rm Na}$ inside and outside
the channels (which is simply $N^{\rm In}_{\rm Na}/V_{chan}$ and
$N^{\rm Out}_{\rm Na}/(L^3-V_{chan})$ where $V_{chan}$ is the volume of the
channels) as a function of $T$. It is immediately noticeable that 
the sodium density inside the channels is
higher than the one outside and in particular at low temperature 
(below $T_g$). So far we knew that the channels are a subspace highly 
visited by the sodium atoms (they have been defined this way) but now we can
conclude in addition that they are also a subspace of \textit{high sodium
density}. We emphasize that a higher sodium concentration does not imply
necessarily a clustering of alkalis~\cite{Greaves,seg}. Actually we 
have shown that no clustering can be observed on a single 
snapshot of the simulation box \cite{Jund}. It is also noticeable that 
the two curves converge towards the limit of uniform density at high 
temperature. This is another evidence that at high temperature the channels 
do not exist and that what we have defined as channels are randomly distributed clusters.\\
Once we know the position of the sodium atoms with respect to the channels
it is naturally of interest to calculate a mean ``residence'' time of 
the Na atoms 
within the channels. This can be done by determining the probability 
$P(0,t)$ of a sodium atom inside the channels at $t = 0$ to 
be inside the channels after a time $t$. $P(0,t)$ is given by:
\begin{eqnarray}
P(0,t)=\frac{1}{N^{\rm In}_{\rm Na}(0)}\sum_{i=1}^{N^{\rm In}_{\rm Na}(0)}n_i(t)
\end{eqnarray}
with:
\begin{eqnarray}
\left\{
\begin{array}{l}
n_i = 0 \text{ outside the channels}\\
n_i = 1 \text{ inside the channels}\nonumber
\end{array}
\right.
\end{eqnarray}
This probability is represented in \textsc{Fig.}~5 at different temperatures.
Since we save the configurations every 1.4~ps (1000 MD steps), the fast 
decay from 1 at shorter times can not be seen in \textsc{Fig.}~5 but 
since the channels have been determined with this ``time step'' we have 
decided to calculate $P(0,t)$ in the same conditions. This means 
therefore that our probability is slightly overestimated since we do not 
take into account the vibrational effects.\\
In \textsc{Fig.}~5 we see that $P(0,t)$ decreases and converges towards 
the long time limit $N^{\rm In}_{\rm Na}/N_{\rm Na}$ very rapidly at
high temperature while at low temperature ($T \leq 2300$~K) this
limit is not reached after 1.5~ns. After subtracting the long time limit, 
these curves can be fitted reasonably well by an exponential function 
with a time
constant $\tau$ which is represented versus 1/T in a linear-log 
plot in the inset of \textsc{Fig.}~5. From this inset it is clear that
$\tau(T)$, the typical residence time, shows an Arrhenius behavior 
with an activation energy around 1.45~eV. This activation energy is 
larger than the activation energy of the Na diffusion 
process and other more specific characteristics of the Na 
displacement (hops between preferential sites, forward-backward jumps, 
decorrelation mechanism) which is around 1.3~eV \cite{Jund}. This seems 
to indicate that the mechanism governing the expulsion of the Na ions 
outside the channels is energetically more costly than the other 
mechanisms controlling the mean displacement of the ions. \\
Finally we have decided to investigate the influence of the location of the
sodium atoms on their diffusion by calculating ``conditional'' mean 
squared displacements (MSD) ($ R^2(t) = \left< \left| \vec{r}(t)-\vec{r}(0) 
\right|^2 \right>$). On the one hand we have calculated $d^2(t)_{\rm In/Out}$ 
which is the ``MSD'' of the atoms inside/outside the channels at 
time $t = 0$ {\em and} $t \neq 0$ (this is a reasonable calculation since
we know that $P(0,t)$ remains non zero over the length of the simulation).  
These quantities are represented in \textsc{Fig.}~6
for (a)~$1700$~K,~(b)~$1900$~K and~(c)~$2300$~K. On the other hand we have
calculated $R^2(t)_{\rm In/Out}$ strictly speaking: in that case the atoms are 
\textit{continuously} inside/outside the channels between $0$ and $t$. In
this latter case
the number of steps an Na atom is consecutively inside/outside the channels 
is small especially at high temperature therefore this quantity is only 
represented for $1700$~K in the inset of \textsc{Fig.}~6~(a). From the 
results represented in \textsc{Fig.}~6 we see that at low temperature 
$(T = $1700,~1900~K) $d^2_{\rm In}$ is always larger  
than $d^2_{\rm Out}$ (at long times the two curves collapse since the 
history of the In and Out atoms becomes the same on average). This 
difference is maximum after $\approx 0.1$~ns when $d^2_{\rm In}$ is 
twice as important than $d^2_{\rm Out}$. This indicates clearly that 
the Na atoms are more mobile inside the channels. This result is coherent 
with the findings of Horbach \textit{et al.}~\cite{horbach}
who state that the ``sodium atoms move quickly between preferential sites'' 
but has never been quantified directly so far. It is confirmed in the inset 
of \textsc{Fig.}~6~(a) by the difference between the MSD of the Na atoms 
that are moving only inside the channels and the MSD of the sodium atoms 
moving only outside the channels. 
At $T=2300$~K (\textsc{Fig}.~6~(c)), we see that the difference 
between $d^2_{\rm In}$ and $d^2_{\rm Out}$ vanishes even though the 
channels still exist. Nevertheless as shown in \textsc{Fig}.~3 
at that temperature the channels are not static anymore and therefore
the distinction between In and Out has no clear meaning. 
At even higher temperatures (not shown), this is of course also true, and
the two curves are superimposed. We can hence conclude 
that when the channels are formed and ``static'' (lowest temperatures), the 
sodium diffusivity is higher when the Na atoms are inside the channels.

In summary, with the use of classical molecular dynamics simulations
on Na$_2$O-4SiO$_2$ systems we have analyzed in more detail the properties
of the sodium trajectories  inside the glassy matrix and especially the 
differences between the Na atoms inside the diffusion channels and those 
outside of these channels (the existence of these channels defined as the
fraction of space mostly visited by the Na atoms has been shown in a previous
study~\cite{Jund}).\\
Firstly we have shown that the average potential energy of the Na atoms 
is the same whether they are In or Out. This is also true concerning
their local environment except a slightly higher Na coordination number
for the sodium atoms inside the channels. This is coherent with the fact
that, in comparison with the rest of the system, the concentration 
of sodium atoms at low temperature (T $\leq$ 3000 K) is higher inside
the channels. At high temperature the matrix can not be considered as frozen
and therefore the channels do not exist anymore. We have defined a
characteristic residence time inside the channels $\tau(T)$ that shows an
Arrhenius dependence with an activation energy around 1.45~eV. This value
is slightly higher than the characteristic energy of the other
activated processes present in this system.
Finally, we have shown that, at low temperature when the channels can be
considered as frozen, the mobility of the sodiums inside the channels is 
larger than the one of the Na atoms in the rest of the system: in that 
sense one can really speak about channel diffusion.\\
In order to find the origin of this effect and in particular the
role of the SiO$_2$ matrix, one might change the vibrational
characteristics of the matrix or those of the cation and see how these changes 
affect the sodium dynamics. These are investigations currently under way.

We thank W. Kob for interesting discussions. Part of the numerical 
calculations were performed at ``Centre Informatique
National de l'Enseignement Sup\'erieur'' in Montpellier, France.

\newpage 

\begin{figure}
\psfig{file=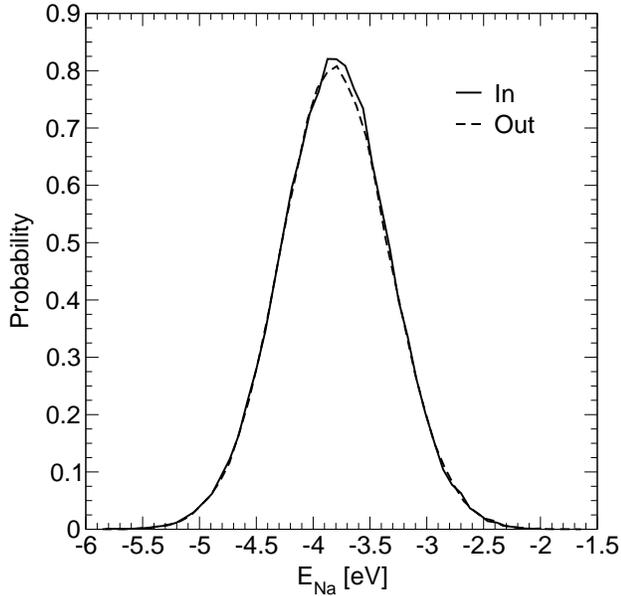,width=8.5cm} 
\caption{ 
Individual potential energy distributions of the sodium atoms
inside and outside the channels (1700~K).}
\label{fig1}
\end{figure}

\begin{figure}
\psfig{file=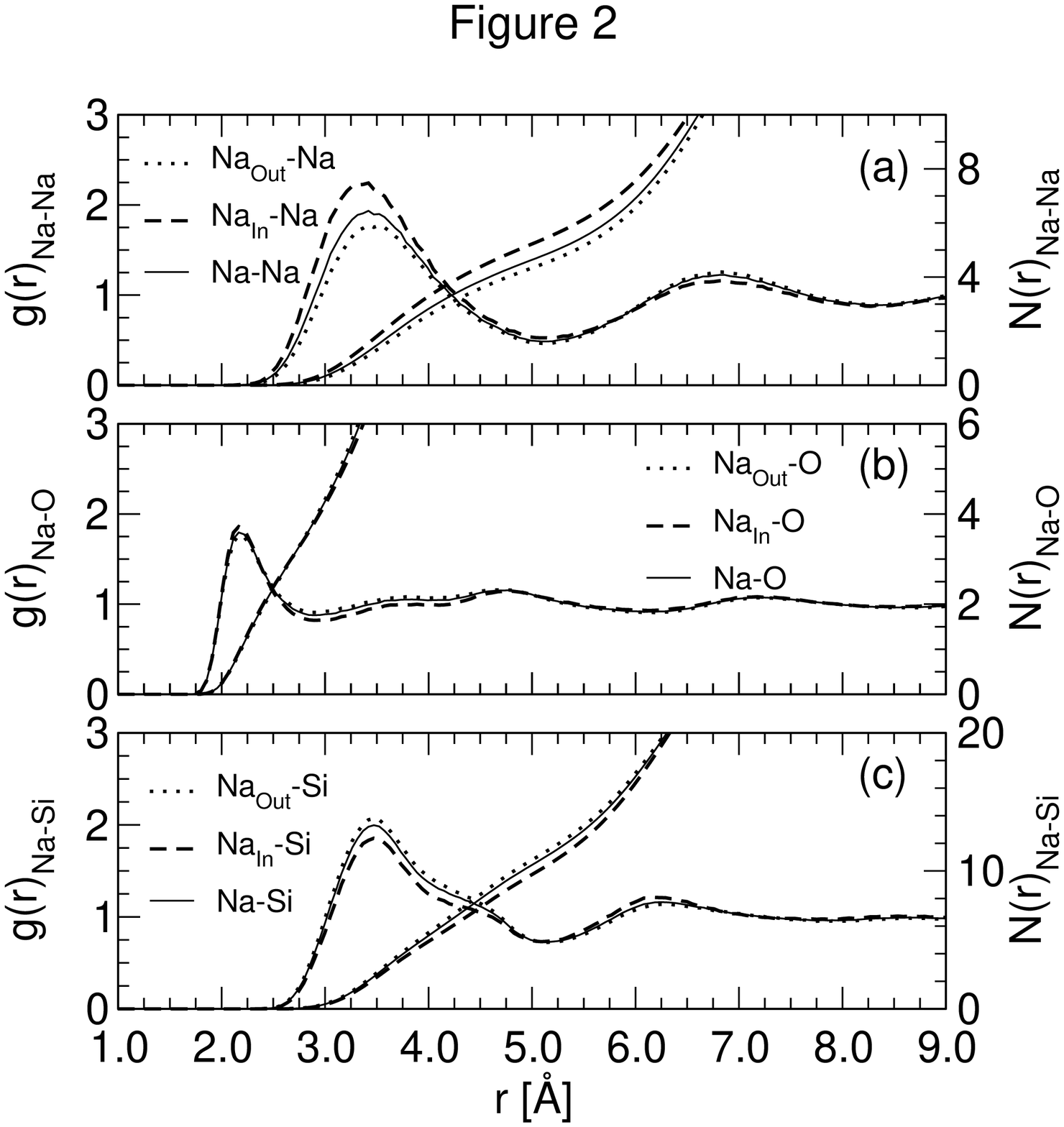,width=8.5cm} 
\caption{
Radial pair distribution functions,~$g(r)$, and integrated
numbers of neighbors,~$N(r), $ at 1700~K for~(a)~Na$_{\rm Out}-$Na, 
Na$_{\rm In}-$Na and Na$-$Na pairs,~(b)~Na$_{\rm Out}-$O, Na$_{\rm In}-$O and
Na$-$O pairs and~(c)~Na$_{\rm Out}-$Si, Na$_{\rm In}-$Si and Na$-$Si pairs.}
\label{fig2}
\end{figure}

\begin{figure}
\psfig{file=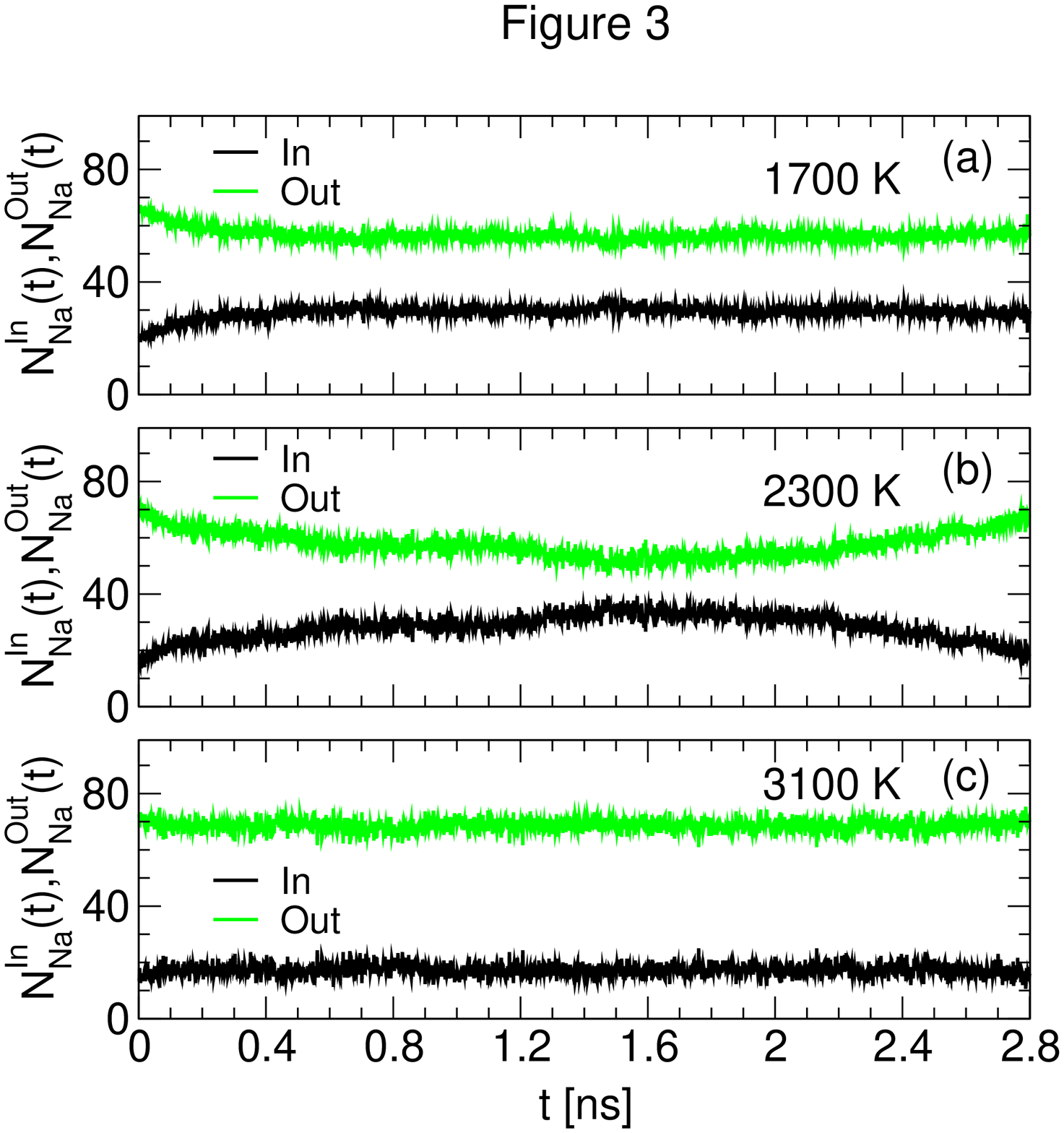,width=8.5cm} 
\caption{
Number of sodium atoms inside and outside the channels versus
time at~(a)~1700~K,~(b)~2300~K~and~(c)~3100~K.}
\label{fig3}
\end{figure}

\begin{figure}
\psfig{file=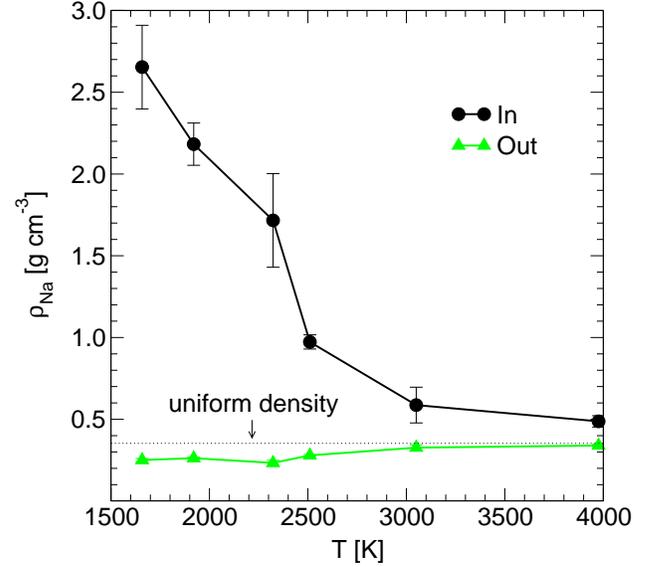,width=8.5cm} 
\caption{
Sodium density $\rho_{\rm Na}$ inside and outside the channels versus 
temperature.}
\label{fig4}
\end{figure}

\begin{figure}
\psfig{file=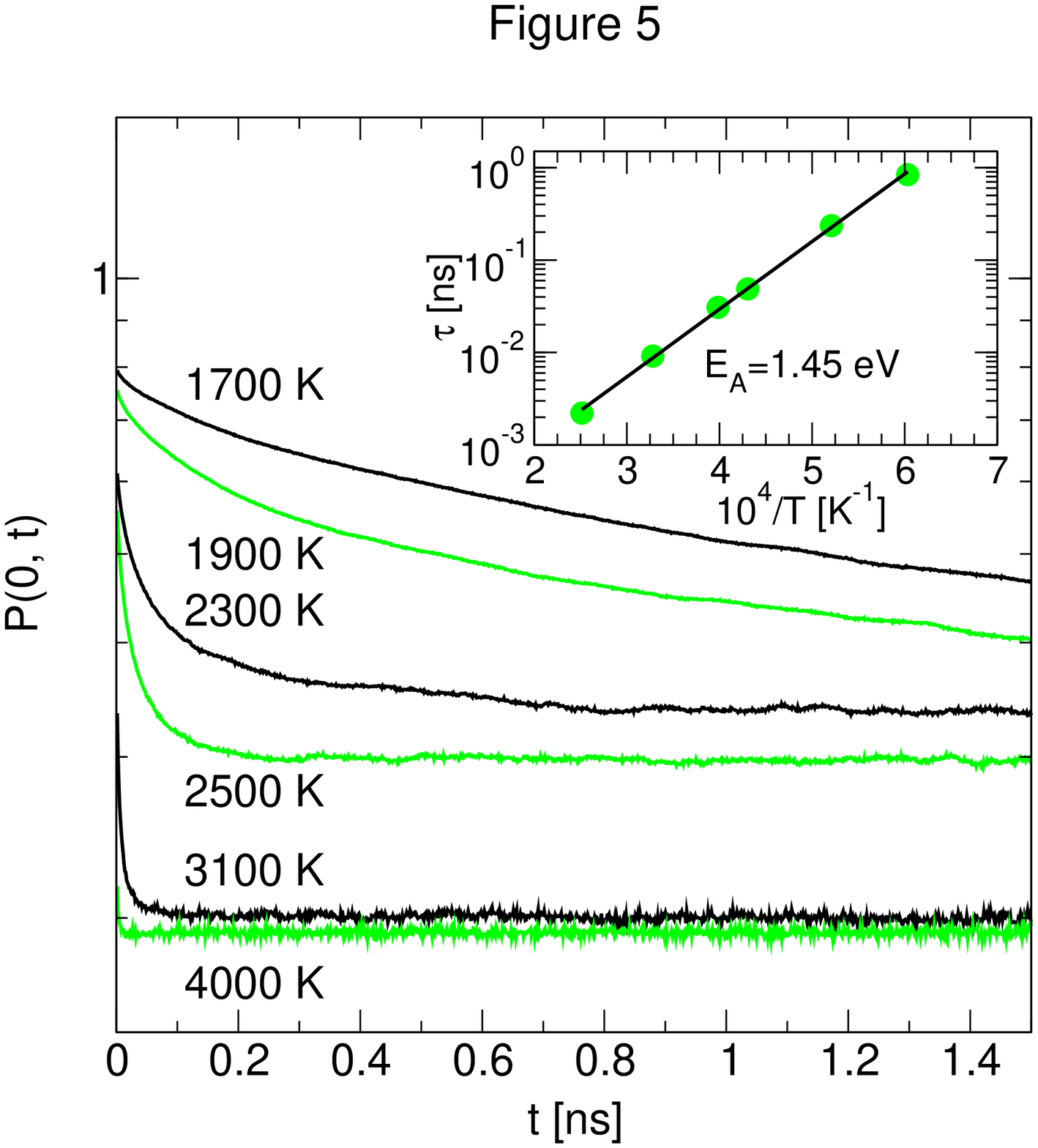,width=8.5cm} 
\caption{
Probability $P(0,t)$ that a sodium atom is inside the channels at time 
$t = 0$ and time $t \neq 0$ for different temperatures. In inset: 
$\tau$ versus 1/T where $\tau$ is the time constant of an exponential 
fit of $P(0,t)$}
\label{fig5}
\end{figure}
\vspace*{-0.8cm}
\begin{figure}
\psfig{file=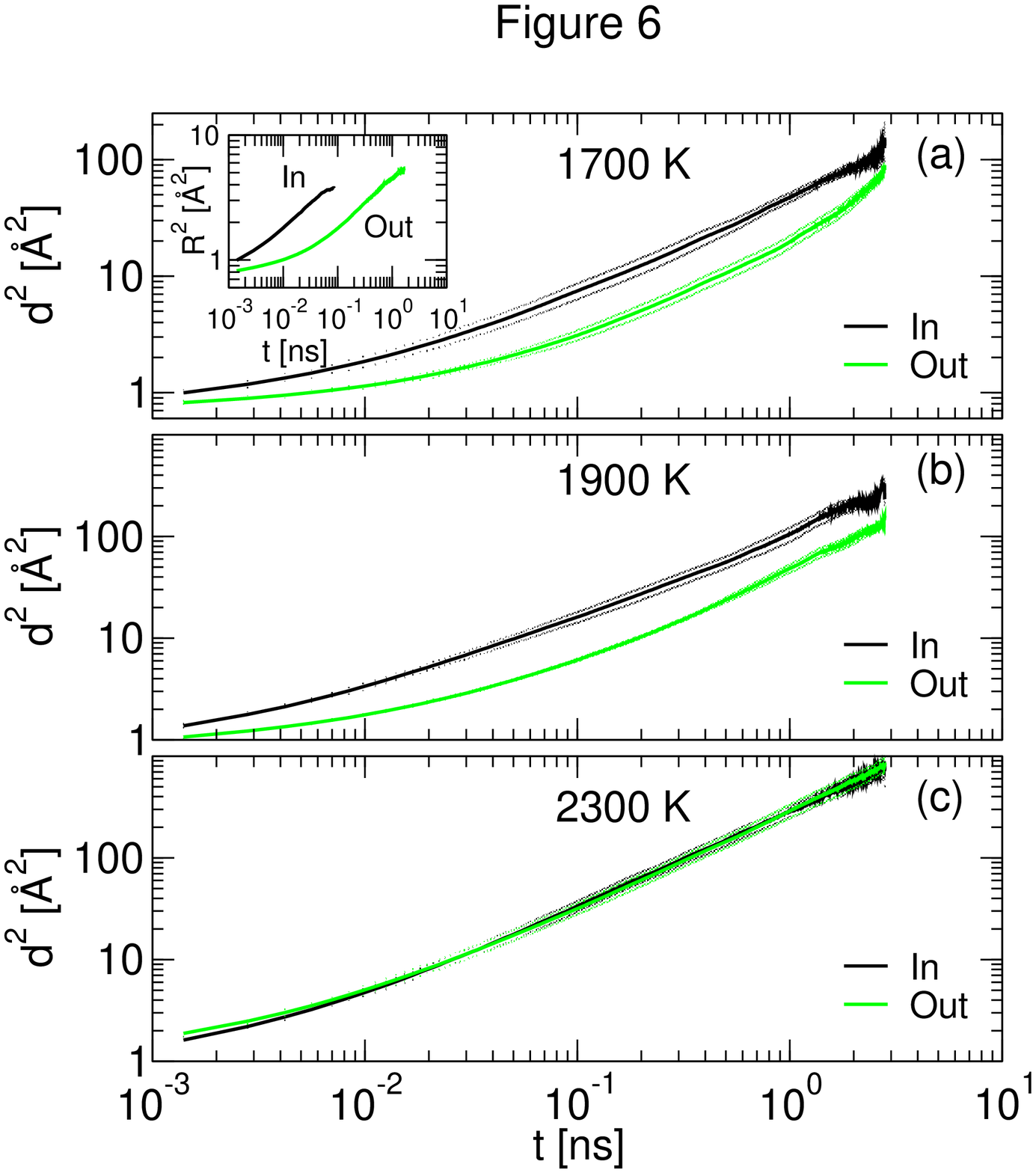,width=8.5cm} 
\caption{
$d^2(t)_{\rm In}$ and $d^2(t)_{\rm Out}$
(see text for definition) at~(a)~$1700$~K,~(b)~$1900$~K and~(c)~$2300$~K.
Inset: Mean square displacements $R^2(t)_{\rm In}$ and $R^2(t)_{\rm Out}$ 
(see text for definition) at 1700~K.}
\label{fig6}
\end{figure}

\end{document}